\documentclass[aps,prl,reprint,nobibnotes]{revtex4-2}
\usepackage[T1]{fontenc}
\usepackage[utf8]{inputenc}
\usepackage{amsmath,amssymb,amsfonts,bm}
\usepackage{graphicx}
\usepackage{float}
\usepackage{hyperref}
\usepackage{comment}
\usepackage{physics}
\usepackage{bbold}
\usepackage{csquotes}
\hypersetup{colorlinks=true,linkcolor=blue,citecolor=blue,urlcolor=blue}

\newcommand{\E}{\mathbb{E}}

\usepackage{xcolor}

\begin{document}
	\title{On the relaxation problem in statistical mechanics}
	\author{Giuseppe Del Vecchio Del Vecchio}
	\email{giuseppedelvecchiodelvecchio@gmail.com}
	\affiliation{Laboratoire de Physique de l'Ecole Normale Sup\'erieure, CNRS, ENS and PSL Universit\'e, Sorbonne Universit\'e, Universit\'e Paris Cit\'e,
		24 rue Lhomond, 75005 Paris, France}
	
	\begin{abstract}
		We reformulate the relaxation problem in statistical mechanics by making explicit what are the \emph{operational} objects subject to relaxation: the local time statistics of the recorded signal $Z(t)$. These local time statistics are simply the estimated histograms of observations $\{Z(t_i)\}_{i=1}^M$ performed at uniformly random times $\{t_i\}_{i=1}^M$ by a clockless observer. The subject of prediction is a belief about a future fresh out-of-sample reading of a measurement outcome whose distribution is inferred from the mathematical model believed to be true. For finite bounded systems of $N\ge 1$ degrees of freedom global irreversible relaxation of predictions can occur but special initial conditions exist. The form
		of the predictions depends on certain loss functions whose choice is up to the particular observer. Finally, entropy is given a learning interpretation as mutual information between the observer and the unknown past of the system under consideration and, in complete generality, its stationary value depends on the information available.
	\end{abstract}
	\maketitle

	\textit{Introduction.} The foundational problem of statistical physics asks: why are physical systems assumed to obey deterministic reversible dynamics described by probability distributions? Are these distribution time-independent? At the most basic level, given the initial conditions, Hamilton's equations are deterministic: where is randomness coming from then? Is the large number of particles $N\gg 1$ necessary to justify probability \cite{gibbs1902, Huang_1987, landau2013statistical, Kardar_2007}? A second conceptual obstacle is that time reversal symmetry enjoyed by dynamics  \cite{LambRoberts1998, Frigg_Werndl_2024} implies, at an ensemble level, that all functions of the energy $P(E)$ are stationary. How is then relaxation possible? Do we need an act of faith and believe the Boltzmann ergodic hypothesis postulating the microcanonical distribution \cite{Sklar_1973, Earman_1996, Moore_2015}? From these facts, irreversibility is then typically seen as a property of macroscopic systems for which $N\gg 1$ \cite{Lebowitz1993BoltzmannsEA, BALDOVIN20251}. Even more dramatically, in quantum systems global relaxation is believed to be impossible owing to unitarity of the Schr\"odinger equation. Recent approaches seem to suggest that relaxation is only true locally in the thermodynamic limit \cite{Cramer_2008, Barthel_2008, Cramer_2010, Sirker_2014, Essler_2016}. Is this actually true?  There seems to be no clear unified answer or interpretation to all such questions \cite{Jaynes_1967, OPenrose_1979, Uffink2006, Mori2018, Frigg_Werndl_2024}.

	It is unquestionable that statistical mechanics is one the pillars of modern physics and its methods have been applied in the most diverse fields like economics \cite{Bouchaud_Potters_2003}, social dynamics \cite{Castellano_2009}, biology \cite{Sella_2005}, network science \cite{Reka_2002}, complex systems \cite{Stanley_1996}, combinatorial optimization \cite{Hartmann_2005}, coding theory \cite{montanari_2007}, machine learning \cite{decelle_2023} and many others. Yet, the unease felt the very first moment we are confronted with the postulates and interpretations of this theory is strong and common to all of us, especially as students. Hence, the suspicion that statistical physics and thermodynamics are not properly understood compared to other theories of physics seems to be well grounded and a proper understanding of its foundations is highly desirable.
	
	In this article we would like to put the role that  inference and information play in a deterministic physical theory on firm grounds. The end result of the discussion will be a hybrid theory where the evolution rule comes from a postulated mathematical model (the Hamilton's equations) and irreversibility from the operation of measurements which force a statistical description. 
	
	\begin{figure}[h!]
		\includegraphics[scale=0.5]{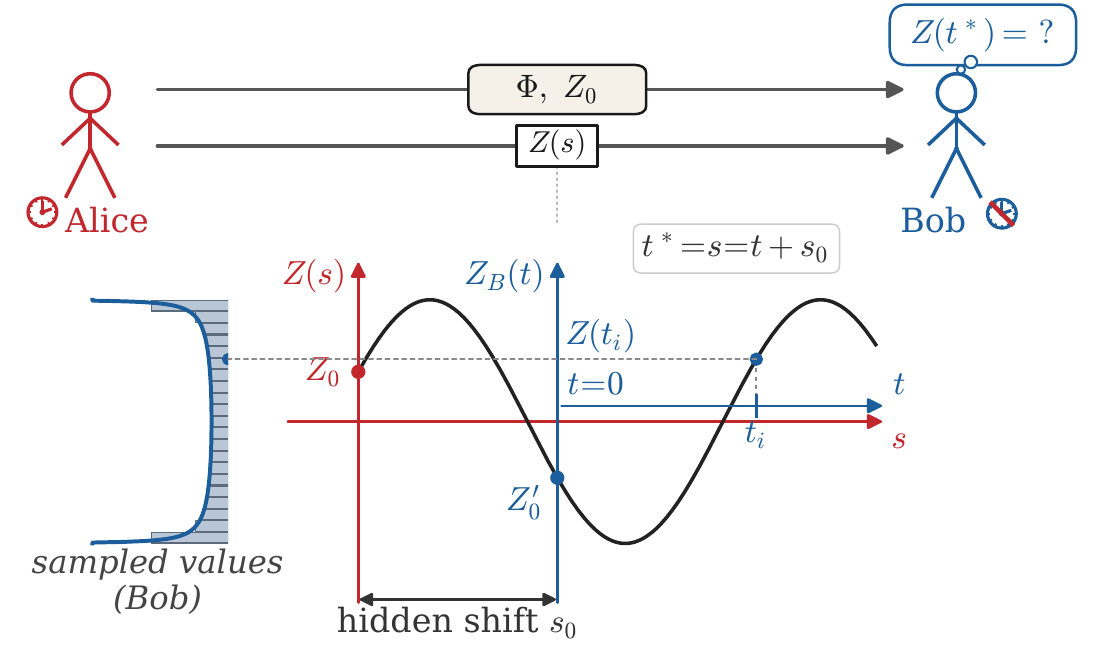}
		\caption{Alice prepares the bounded signal moving as $Z(s) = \Phi_s(Z_0)$. Bob receives the signal after $s_0>0$. Bob is uncertain about the future because the signal has value $Z_0'\neq Z_0$ at Bob's time $t=0$ corresponding to $s=s_0$. Bob is clockless, i.e., he is interested in the ordinates of the signal and his task is to guess a value $Z(t^*)$ in the arbitrary future $t^*>0$ (see Eq.~\eqref{eq:bob_future}).  Using knowledge of $\Phi$ and $Z_0$, Bob then builds the histogram on the left and computes probability law using the samples $\{Z(t_i)\}_{i=1}^M$. This leads to the estimate Eq.~\eqref{eq:stat_pred}.}
		\label{fig:shift}
	\end{figure}

	Our point of view is of course very close to Jaynes  who was certainly one of the pioneers to bring inference and physics together \cite{Jaynes1957a,Jaynes1957b,Jaynes_1967,Jaynes1968}. Jaynes maximum entropy approach imposes macroscopic constraints such as average energy and other conservation laws to derive the least committal probability distribution \emph{at a given unspecified time}. Nevertheless, while from a computational point of view maximum entropy methods reproduce the prescriptions of statistical mechanics, it is true that they do not give a dynamical justification of the theory \cite{BALDOVIN20251}. 
	
	Our contribution will be precisely to show that by considering as prior information the whole data specifying the dynamical model believed to describe a certain physical system leads to a useful conceptual improvement. The estimated probabilities are objective, i.e., computable frequencies, once the prior information is fixed but subjective in the sense that they change as the priors change for different observers. 
	To support our view, we notice that, besides the traditional works of Jaynes, recent works on entropic dynamics demonstrate that inference constitutes a powerful principle in physics, generalizing classical `actions' \cite{caticha2007, caticha2011, Caticha2015}. Here, even quantum mechanics is reinterpreted and derived from an inference principle.

	The basic fact that we wish to consider seriously - which is lacking in Jaynes formulation - is that before any type of measurement an observer does not know the outcome \cite{peres1978, Johnson_2012}.  Why would an observer need a measurement otherwise?
	In other words, uncertainty becomes certainty only after a measurement \cite{Shannon_1948}. A prediction is necessary only before the next, still to be seen, measurement outcome and represents a belief. These beliefs can be updated using the rules of inference \cite{Jaynes_2002} which use previous information, collected via measurements, to guess new possible readings.
	
	That the operation of measurements through time averages is relevant in statistical mechanics is discussed in standard books \cite{kh49,Huang_1987, Parisi_1998, Frigg_Werndl_2024, BALDOVIN20251}, where that average is justified by the apparatus operating slowly compared to the dynamics. But why a plain time average and not a weighted one? And even granting the argument, the validity of statistical mechanics still hinges on an ergodic theorem, hard to establish in general \cite{Sklar_1973, Earman_1996, Moore_2015}.
	
	\textit{Ideal clockless measurements.} To see what the role of measurements is we can imagine a `clockless' observer that collects samples $\{Z(t_i)\}_{i=1}^M$ from a deterministic function of time $Z(t)$. 
	In particular, given the path $(t,Z(t))$ the operation of collecting a sample at some time $t_i$ `without looking at the clock' can be written mathematically as $\pi((t_i, Z(t_i)))=Z(t_i)$.  We may call the projection $\pi$ the measurement operator. In particular, the nature of the dynamics does not really matter for the statistical properties of the dataset $\{Z(t_i)\}_{i=1}^M$ to be well defined. 
	
	To stay as close as possible to the original statistical mechanics formulation and investigate the problem of its foundations we will focus on \emph{bounded} systems that are \emph{assumed} to obey time reversal invariant and autonomous \emph{deterministic} dynamics. In formulas this is expressed by a rule $\Phi_{t}$ mapping $Z(t_0)$ to $Z(t_0+t)$ with the property of a group w.r.t. $t$ \cite{goldstein1980}. Time reversal symmetry is the statement that there is an involution $R^2 = 1$ such that $R\circ \Phi_t \circ R = \Phi_{-t}$. As boundedness implies a finite number (or volume) of possible microscopic states, if the IC is $Z_0$ and if $Z(t_i)=\Phi_{t_i}(Z_0)$ is a sample collected at time $t_i$, stationarity of the samples statistics seems a-priori very plausible: the signal cannot go beyond its limits and must come back remaining confined. And since prototypical systems from which statistical mechanics originated are bounded, like a gas in a box, we will restrict to this case. 
	
	From these considerations it follows that, as long as the dataset $\{Z(t_i)\}_{i=1}^M$ is concerned, not even time reversal symmetry - a property of the path $(t,Z(t))$ not of the ordinate $Z(t)$ alone \cite{LambRoberts1998} - seems to be an obstacle for stationarity or `irreversibility'. See Fig.~\ref{fig:shift}. Indeed, the measurement operation $\pi$ defined above loses the ordering information of the samples with respect to (w.r.t.) times $t_i$ and distinguishing past from future becomes impossible.

	To clarify the role played by  inference we have found useful to think in terms of a timely branch of statistics called learning theory \cite{Vapnik1999,GneitingRaftery2007, Hastie2009, MEHTA20191}. This framework of ideas has demonstrated enormous success in recent times when applied to machine learning problems.
	Here, the property that is asked to a given statistical model supposed to represent reality is that of \emph{generalization}, a term borrowed from psychology \cite{Sephard_1987}. Generalization means that a particular model must perform well when tested on new examples not present in the dataset used for training. The out-of-sample error is called generalization error \cite{MEHTA20191}. The minimization of this error provides the inference rule which is specific to the task that the statistical model is supposed to perform \cite{Hastie2009, MEHTA20191}.

	In the same way, in this work we will ask: 
	\begin{displayquote}
		\textit{Given the deterministic rule $\Phi$ and an initial condition (IC) $Z_0$, what is our prediction for a new and unseen measurement outcome at some future time?}
	\end{displayquote} 
	The task dependence will be the prediction about some property of particular observable or class of observables at some future time. Predictions can be point estimates or whole probability distributions as we will show below.

	\textit{Alice and Bob.} To understand our formulation it is best to think about the following situation: let Alice be in possession of a  clock and let $s$ be her time coordinate. Alice prepares the system moving with a certain deterministic  dynamical rule $\Phi_s$ starting from a certain IC $Z_0$ at initial time $s=0$ that she records from her clock. For Alice the signal at time $s$ is 
	\begin{equation}\label{eq:alice_Z}
		\text{Alice:} \quad Z(s)= \Phi_{s}(Z_0)
	\end{equation}
	and the signal $Z(s)$ is perfectly determined at any $s> 0$ (Alice's future). 
	She then puts the system in a closed box and hands it in to Bob at time $s_0>0$ (in her coordinates). At the same $s_0$ she communicates to Bob the following information: i) the precise form of the dynamical rule $\Phi$ and ii) the precise value of the IC $Z_0$. 
	
	\begin{figure}[h]
		\includegraphics[scale=0.8]{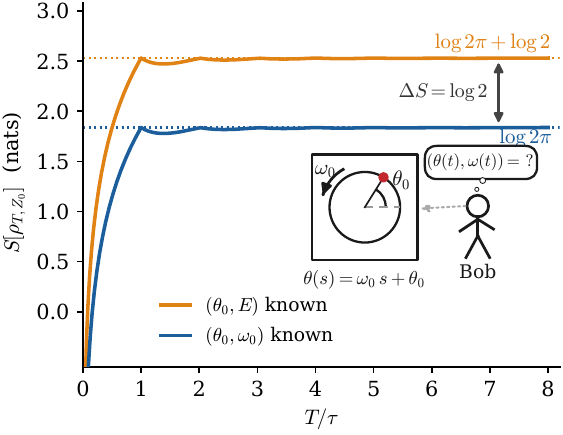}
		\caption{Optimal generalization error $L^*_T = S[\rho_{T,Z_0}]$ in Eq.~\eqref{eq:ent_main} estimated by sampling, for the joint recorded state $Z=(\theta,\omega)$ on the unit ring. Blue: known $(\theta_0,\omega_0)$; orange: known $(\theta_0,E)$. The ordinate is the regularized global entropy: angular bins add $-\log\Delta\theta$, while a known conserved velocity occupies one bin and adds zero \cite{supp}. The relation between forward and backward sampling under time reversal is discussed in \cite{supp}.}
		\label{fig:gen_error}
	\end{figure}

	Point i), i.e., the knowledge of the dynamical rule $\Phi$ is what an established physical theory represents, the Hamilton's equations for example: we have no doubt about their validity. As for point ii), we allow $Z_0$ a certain variability. Indeed, to justify a statistical description it is typically assumed that the source of randomness in a large system is the lack of knowledge of $Z_0$ leading to ensemble descriptions `a la Gibbs' \cite{Mori2018, BALDOVIN20251}. It is our intent here to show that while in most experiments this is certainly true, it is not the only possibility when observing a certain phenomenon.
	Indeed, the literature already distinguishes very well between fixed (quenched) \cite{Calabrese_2007, Essler_2016, DAlessio2016} and random (annealed)  \cite{spohn2012large, Delvecchio_2020, Chakraborti_2022} IC. Yet the precise relations of these two situations w.r.t. the relaxation problem is not clear, at least to us.

	Now let us assume Bob does not have a physical clock to read time but has a \emph{notion} of time and uses the same units as Alice. Let us call Bob's time coordinate $t$ and fix his origin $t=0$ at the moment he receives the box from Alice, i.e., $s_0$. This $t$ is the time coordinate Bob would use if he had access to a copy of Alice's physical clock yet not synchronized with it. 
	The main point is that, even knowing $\Phi$ and $Z_0$, Bob is not in a position to calculate the future values $Z_B(t)$ in his coordinate system $t$. Indeed, when he receives the system from Alice at $t=0$, the true value of the signal is, generally speaking, $\Phi_{s_0}(Z_0)=Z'_0\neq Z_0$. Lacking knowledge about $s_0$ Bob does not know $Z_0'$. 
	Applying a time shift Bob finds that his time coordinate $t$ is related to Alice's time coordinate $s$ by $s = t+s_0$. See Fig.~\ref{fig:shift}. Using this result in Eq.~\eqref{eq:alice_Z} he finds  
	\begin{equation}
		\label{eq:bob2}
		\text{Bob:}\quad Z(s) = \Phi_{t+s_0}(Z_0)\equiv Z_B(t)\,.
	\end{equation}
	The interpretation of Eq.~\eqref{eq:bob2} is the following: on left hand side (l.h.s.) there is the present (true) value $Z(s)$ in Alice's coordinates, which Alice knows perfectly by Eq.~\eqref{eq:alice_Z}; on the right hand side (r.h.s.) there is Bob's present $Z_B(t) = \Phi_{t+s_0}(Z_0)$ which is uncertain to him because he does not know the shift $s_0$. 
	Bob's uncertainty comes from the hidden shift $s_0$ between the two coordinate systems $s$ and $t$. Thus, we can set
	\begin{equation}\label{eq:bob_future}
		t^* \equiv s= t+s_0
	\end{equation}
	where $t^*$ is the future in Bob's coordinates, $t$ the present, $s_0$ the hidden origin and $s$ Alice's present.

	As we already mentioned, the fact that Bob is clockless is the feature of any observer that is only interested in measurement readings producing the value of $Z(t^*)$ at some $t^*$ in the future but not to the reading of $t^*$. Anyway, even if Bob could record $(t,Z(t))$ using a physical copy of Alice's clock, the very fact that the two are not synchronized, i.e., that Bob ignores the time at which the evolution started, makes the future uncertain. This non-synchronization is what happens in most scientific enquiries where the observer did not prepare the system herself. Clearly, had Bob been in possession of a clock he could measure at $t=0$, find $Z_0'$ and compute $Z_B(t) = \Phi_t(Z_0')$ from the knowledge of $\Phi$. Yet, before the very first measurement $Z_0'$ will be uncertain.

	\textit{Sampling.} 
	Due to uncertainty, Bob's task is to have a statistical prediction for $Z(t^*)$ at any \emph{arbitrary} future time $t^*$. How can he make such a guess using all the information he has?
	
	Bob can ask a simple practical question similar to the one we quoted in the Introduction: ``what histogram would I find if I had physically measured the system at some \emph{future times} $\{t_i\}_{i=1}^M$ in an observation window $[0,T]$ given $\Phi$ and $Z_0$?'' To answer that we notice that since Bob has chosen his time origin at $t=0$ in Eq.~\eqref{eq:bob_future} and since he does not have a clock, these measurement times in the future are i.i.d. uniformly distributed in $[0,T]$ because of the unknown time shift $s_0$ (this is also a maximum entropy assignment to $s_0$). Said in other words, this is because Bob has no information distinguishing any time in $[0,T]$. 
	
	To make maximal use of the information about $\Phi$ and $Z_0$ Bob \emph{imagines} making a fresh measurement of the signal and getting $Z(t_i)=\Phi_{t_i}(Z_0)$ at time $t_i$. He can do that, for example on a computer, without measuring the actual physical system received from Alice because he knows both $\Phi$ and $Z_0$. As Bob took the origin at $t=0$, which is anyway an arbitrary choice, by virtue of Eq.~\eqref{eq:bob_future} this sampling procedure can be interpreted by Bob as receiving $M$ independent systems for which the preparation shift in Fig.~\ref{fig:shift} is $s_0=t_i$ for $i=1,\dots, M$. Importantly, all the preparations share the same $Z_0$ and the same $\Phi$.

	Continuing the sampling described above, Bob collects the dataset $\{Z(t_i)=\Phi_{t_i}(Z_0)\}_{i=1}^M$ where each sample  is an i.i.d. random variable because i) the $t_i$'s are i.i.d. and ii) the rule $\Phi$ is deterministic and so does not introduce temporal correlations between the samples. He then constructs the empirical measure
	\begin{equation}\label{eq:emp_freq}
		\hat \mu(A)=\frac{1}{M}\sum_{i=1}^M \mathbb{1}(Z(t_i)\in A)\,.
	\end{equation}
	Here $A$ can be though of as a bin of size $|A|$. The r.h.s. is a counting statistics, well known in physics, see Refs. in \cite{Burenev_2024}. The random variables $\mathbb{1}(Z(t_i) \in A)$ in Eq.~\eqref{eq:emp_freq} are i.i.d. Bernoulli variables with mean $\mu_{T,Z_0}(A)=T^{-1}\int_0^T \mathbb{1}(Z(t)\in A)\dd t$ and variance $\mu_{T,Z_0}(A)(1-\mu_{T,Z_0}(A))$. Hence, the variance w.r.t. the $\{t_i\}_{i=1}^M$ of $\hat \mu(A)$ is $O(M^{-1})$ and, in the ideal limit $M=\infty$, the law of large numbers holds. Thus, Bob obtains an estimate for the probability 
	\begin{equation}\label{eq:stat_pred}
		\Pr(Z(t^*)\in A\,|\mathcal{I}) = T^{-1} \int_{0}^{T} \mathbb{1}(\Phi_t(Z_0)\in A)\dd t
	\end{equation}
	where  $\mathcal{I}\equiv(t^*\in[0,T], \Phi, Z_0)$ is the conditioning information set known to Bob and where we used the fact that Bob knows the dynamical rule $\Phi$ so that $Z(t)=\Phi_t(Z_0)$. This conditioning information set $\mathcal{I}$ in Eq.~\eqref{eq:stat_pred} deserves to be emphasised as the estimated probability is conditional on $\mathcal{I}$: changing $\mathcal{I}$ changes the predicted probability. Notice also how the hidden shift $s_0$ in Eq.~\eqref{eq:bob2} plays a marginal role in this estimate and its only effect is only to make the $\{t_i\}_{i=1}^M$ i.i.d. from the point of view of Bob. We also notice that the r.h.s. of Eq.~\eqref{eq:stat_pred} is well known in the theory of stochastic processes as occupation time measure \cite{Levy1940, feller1950, Majumdar2005, majumdar2006} and it is the familiar time average appearing in discussions about the justifications of statistical mechanics \cite{Huang_1987,  Kardar_2007, landau2013statistical}. Here it only appears because of Bob's uncertainty about the past and the use of knowledge of $\Phi$ and $Z_0$ that he makes to enquire about the system's future. 
	
	We stress that although Eq.~\eqref{eq:stat_pred} is Bob's best guess given his information, he can \emph{eventually} compare these epistemic frequencies with those recorded by a physical device built to count the same events. Equation~\eqref{eq:stat_pred} is thus a belief before measurement that becomes objectively right or wrong after it. Disagreement means either Bob's prior information $\mathcal{I}$ was insufficient or the device was not built for this purpose. Having clarified this important point, we will now focus on a human Bob whose task is to make a guess given the prior information.
	
	\textit{Stationary prediction.}  Since Bob wants a prediction for $Z(t^*)$ for arbitrary $t^*$ in the future then he \emph{intentionally} takes $T\to \infty$ in Eq.~\eqref{eq:stat_pred}. He takes this limit just because he is interested in getting a probability that works for arbitrary future times, i.e.,  $t^*\in[0,\infty]$ (recall Eq.~\eqref{eq:stat_pred}). Hence, in this interpretation, it is not the system that is relaxing but Bob that deliberately takes $T\to \infty$ in order to have a probability that works for any future instant $t^*$. 
	
	For now let us comment on that the system's details enters through the bounded dynamics $\Phi$ in that, for each fixed $Z_0$, the limit of Eq.~\eqref{eq:stat_pred}
	\begin{equation}\label{eq:bob_stat}
		\mu_{Z_0}^{\rm st}(A) = \lim_{T\to \infty}\frac{1}{T}\int_0^T \mathbb{1}(\Phi_t(Z_0)\in A)\dd t
	\end{equation}
	may exist or not. If it does, Bob can report and use a stationary prediction using $\mu_{Z_0}^{\rm st}$. Whether this is eventually microcanonical, canonical or not depends on $Z_0$ and on the precise form of the rule $\Phi$, including all the values of all geometrical parameters and eventual scaling limits.  For Hamiltonian systems, the celebrated KAM tori at low energies \cite{BALDOVIN20251} provide explicit examples on the role of the IC. A simpler one is discussed below.
	
	Importantly, formula Eq.~\eqref{eq:bob_stat} is a result of Bob inference that from the knowledge of $\Phi$ and $Z_0$ wants to have a prediction for $Z(t^*)$ with $t^*$ arbitrary in the future. When the limit in Eq.~\eqref{eq:bob_stat} does not exist, Bob will need to keep $T$ finite and use Eq.~\eqref{eq:stat_pred}. 
	Two well known examples of non-existence of the limit in  Eq.~\eqref{eq:bob_stat} are attracting heteroclinic cycles \cite{Gaunesdorfer_1992} and symbolic dynamics generated by horseshoes near transverse homoclinic orbits \cite{Smale_1967, HOLMES1990137}. On the other hand, an exceptionally simple case where the limit in Eq.~\eqref{eq:bob_stat} always exists is that of a dynamics that is reversible and discrete (in both state space and time): here $\Phi$ is a permutation and the limit Eq.~\eqref{eq:bob_stat} converges to the uniform average on the cycle selected by the IC $Z_0$. Thus, in this case, whether Bob can get the `correct' stationary law depends on whether he knows $Z_0$ or not. Indeed, two different $Z_0$ selecting two different cycles (ergodic components, see EM) lead to two different stationary predictions.  In any case, this stationary law describes only what Bob expects for future outcomes not what the actual system is doing in Alice's box. 
	
	Should a physical device record frequencies agreeing with Bob's stationary prediction in Eq.~\eqref{eq:bob_stat}, then the pair $(\Phi,Z_0)$, together with the limit $T\to \infty$, can be considered a good model for the experiment; should they disagree, then either that information was insufficient or the device was not built to record the relevant frequencies for such long times. 
	
	Finally, we note that from Eq.~\eqref{eq:bob_stat} the statistics of arbitrary observables $f(Z)$ is found by push-forward or marginalization $\mu_{f,Z_0}^{\rm st}(\{f(Z)\in A\}) = \mu_{Z_0}^{\rm st}(f^{-1}(A))$, see \cite{supp} for a discussion on the consequences of this global relaxation. This procedure allows, in principle, computation of moments, cumulants and correlation functions. 
	In what follows we will set
	\begin{equation}
		\label{eq:bob_choice2}
		\mu_{\rm st} \equiv \mu_{Z_0}^{\rm st}
	\end{equation}
	where $\mu_{\rm st}$ is the unique distribution supported on a particular ergodic component selected by $Z_0$ (see EM) via the limit Eq.~\eqref{eq:bob_stat}, which Bob can calculate from $\Phi$ and $Z_0$.

	\textit{Generalization error.}
	How large are Bob's average mistakes about the future? To see this recall that $\mu_{T,Z_0}(A)=T^{-1}\int_0^T \mathbb{1}(\Phi_t(Z_0)\in A)\dd t$ is the estimated Bob's measure in the r.h.s. of Eq.~\eqref{eq:stat_pred}. Let also $\E_{\mu_{T, Z_0}}$ the expectation w.r.t. this measure.
	
	For simplicity, let us consider the distribution of the full signal $Z(t)$ as in Eq.~\eqref{eq:emp_freq}. We assume that $Z$ is continuous and that the probability measure in Eq.~\eqref{eq:stat_pred} or Eq.~\eqref{eq:bob_stat} has a density, $\mu_{T,Z_0}(\dd z)=\rho_{T,Z_0}(z)\dd z$. The case where $\mu_{T,Z_0}$ has singular parts is treated in \cite{supp}.
	Now, let $\rho_{\rm p}$ be any probability density that Bob would use to predict that $Z(t^*)\in A$ at some future time $t^*\in[0,T]$ without knowing or using $\Phi$ and $Z_0$. A common loss function in this case is the average negative log-likelihood also known as log-loss \cite{GneitingRaftery2007}. The generalization error in this case is a functional of $\rho_{\rm p}$ and it is given by $L_{T,Z_0}[\rho_{\rm p}]=-\E_{\rho_{T,Z_0}}[\log \rho_{\rm p}(Z)]$, i.e., the expected surprisal. Other loss functions are possible making the predictions observer dependent but here we focus on this illustrative case \cite{DelVecchio_future}. It is simple to see that this functional can be rewritten as \cite{KullbackLeibler1951}
	\begin{equation}\label{eq:kl_dec}
		L_{T,Z_0}[\rho_{\rm p}] = S[\rho_{T,Z_0}] + D_{\rm KL}(\rho_{T,Z_0}||\rho_{\rm p})
	\end{equation}
	where $S[\rho]=-\E_{\rho}[\log \rho(Z)]$ is the Shannon entropy and $D_{\rm KL}(\rho||\sigma)=\E_{\rho} \log(\rho(Z)/\sigma(Z))$ is the KL divergence quantifying the distance between the predictor $\rho_{\rm p}$ and the data distribution $\rho_{T,Z_0}$. Since $D_{\rm KL}(\rho||\sigma)\ge 0$ for all $\rho,\sigma$ \cite{KullbackLeibler1951, mezard_montanari2009, MEHTA20191}, the minimum generalization error is obtained by minimizing $D_{\rm KL}(\rho_{T,Z_0}||\rho_{\rm p})$ (the generalization gap \cite{GneitingRaftery2007}) w.r.t. $\rho_{\rm p}$. For Eq.~\eqref{eq:kl_dec}, the optimal solution is  $\rho^*_{\rm p} =\rho_{T,Z_0}$ \cite{GneitingRaftery2007}. Hence, the optimal generalization error given by
	\begin{equation}
		\label{eq:opt_gen_err}
		L_{T,Z_0}^{*} = S[\rho_{T,Z_0}]
	\end{equation}
	which is the Shannon entropy of $\rho_{T,Z_0}$, which can be computed by Bob knowing $\Phi$ and $Z_0$ as in Eq.~\eqref{eq:stat_pred}.
	As already mentioned above, once the full inferred probability law $\mu_{T,Z_0}$ converges to $\mu_{\rm st}$, all its marginals and all bounded expectations converge to their stationary values. At finite measurement resolution the entropy converges as well \cite{supp}. 
	
	Indeed, as Bob takes $T\to \infty$, the optimal generalization error saturates, possibly non-monotonically, as $L^*_{T,Z_0}\to S[\rho_{\rm st}]$. We further show in EM that Eq.~\eqref{eq:opt_gen_err} is equal to the mutual information between the signal and the initial time shift $s_0$ at which Alice prepared the system and which Bob ignores $I(\Phi_{s_0}(Z_0),s_0)$. Hence, entropy increase is interpreted here as learning about the past of the system up to the maximum  value allowed by the information available and it is not a property of the system rather of the observer, an interpretation which is widely different from the tradition \cite{OPenrose_1979, Lebowitz1993BoltzmannsEA, spohn2012large}.
	
	In Fig.~\ref{fig:gen_error} we report the dynamics of $L^*_{T,Z_0}$ as an example of Bob learning the joint distribution of the angle and the angular velocity $(\theta, \omega)$ of a single particle rotating on a ring of radius $R=1$ when the information about the initial conditions changes. 
	In the first case we fix $Z_0=(\theta_0,\omega_0)$ while in the second case we fix $(\theta_0,E)$ with $E=\frac{1}{2}m \omega_0^2$ being the total energy leaving ${\rm sign}(\omega_0)$ unknown producing $1$ bit $\approx 0.693$ nats of difference. A simple calculation shows that, after regularization \cite{supp}, Eq.~\eqref{eq:opt_gen_err} becomes
	\begin{align}\label{eq:ent_main}
		S[\rho_{T,Z_0}]=&\log(2\pi)-r\,\frac{n+1}{n+r}\log\frac{n+1}{n+r}\nonumber\\
		&-(1-r)\,\frac{n}{n+r}\log\frac{n}{n+r} + \chi
	\end{align}
	where $\chi=0$ if $(\theta_0,\omega_0)$ is known while $\chi=\log(2)$ when only $(\theta_0,E)$ is known. In Eq.~\eqref{eq:ent_main} we defined $n = \left\lfloor\frac{T}{\tau}\right\rfloor$ and $ r=\left\{\frac{T}{\tau}\right\}$ with $\tau = 2\pi /\omega_0$ being the period. From Fig.~\ref{fig:gen_error}, we can see that the entropy plateau is a function of the prior knowledge $\mathcal{I}$ in Eq.~\eqref{eq:stat_pred} through $\chi$ in Eq.~\eqref{eq:ent_main} and, as recalled in EM and shown explicitly in \cite{supp}, only in the second case coincides with the microcanonical Boltzmann entropy calculated from $\rho_{\rm mc}(\theta, \omega) \propto \delta(\frac{1}{2} m \omega_0^2 - H)$. Time reversal relates forward and backward predictive distributions, but does not in general make them identical for the same IC; the precise relation, the role of the measurement bins, and special orbits selected by special ICs are discussed in \cite{supp}.

	Before closing, we remark that what we have shown is that for the special dynamical rules $\Phi$ with the properties considered in this work, as a matter of principle and once measurements are taken into account, neither the number of particles $N$ needs to be large nor special properties beyond boundedness of the dynamics are important to do statistical mechanics with stationary distributions. Irreversible behavior of the estimated distributions can occur, even globally, except for very specific IC \cite{supp}. Comparison of predictions with experiment allows only to assess the validity of the assumed prior information with respect that particular experiment and prediction task and deliberate induction leads to inhevitable difficulties. 
	
	Finally, a large number of particles $N\gg 1$ becomes important only if one wishes to recover thermodynamics relations about average energy and heat. These happen to be linear statistics with a specific functional form \cite{kh49}. See the qualitative discussion in EM. Nevertheless, the issue is delicate and the properties of the IC are still important in the sense that the predicted distributions may or may not be sharp due to $N\gg 1$ and their form need not be of any a-priori specific form: there are infinitely many distributions with the same low order moments. 
	These issues are the subject of a future work \cite{DelVecchio_future}.
	
	\begin{center}
		{\bf Acknowledgments}
	\end{center}
	The author is supported by ANR grant no. ANR-23-CE30-0020-01
	EDIPS. This work was completed during the program Advances in Non-equilibrium physics hosted by Kavli Institute of Theoretical Physics in Santa Barbara, CA. The author benefitted from multiple discussions with various participants during his stay. In particular he would like to acknowledge discussions with S. N. Majumdar, S. Sabhapandit, M. Biroli and G. Mussardo.

	\bibliography{bibliography}

	\clearpage
	\appendix
	\begin{center}
		{\bf End Matter}
	\end{center}
	\section{Unknown $Z_0$ and ergodic decomposition}
	We briefly recall here how the ergodic decomposition of a dynamical system works. Since, at least in this paper, the signal $Z(t)$ is assumed to be bounded there are in general many `ergodic components' in which the system can be found moving. These are simply subsets of the state space such that once $Z(t)$ enters in one of them at some time, it never leaves. 
	
	More precisely, the state space can be partitioned as $\Omega= \cup_\alpha \Omega_\alpha$ where the index $\alpha$ can be continuous or discrete and for $\alpha\neq \beta$ the components satisfy $\Omega_\alpha  \cap \Omega_\beta =\emptyset$. Hence, once the IC $Z_0\in \Omega_{\alpha}$ for some $\alpha$ then $\Phi_t(Z_0)\in \Omega_\alpha$ for all $t\ge 0$. It follows that for reversible laws $\Phi$, for each $Z_0$ there is a unique ergodic component $\alpha_0\equiv\alpha(Z_0)$ that is selected at the beginning of the evolution. 
	The ergodic components $\Omega_\alpha$ might even be very low dimensional subsets of $\Omega$ like the minima of the potential energy or deep wells of a rough potential landscape like in spin glasses \cite{EdwardsAnderson1975}.
	For each IC $Z_0$ in a particular ergodic component $\Omega_\alpha$, the $T\to \infty$ limit of the time average in Eq.~\eqref{eq:stat_pred} gives, when it exists, a unique measure $\mu_\alpha$ which depends only on the label $\alpha$ not on the particular $Z_0$. 
	
	Now, assume Bob does not know the IC $Z_0$ and still needs to estimate the probability distribution of $Z(t^*)$ for future values of measurements. Then he needs a rule to assign a probability to each of them. In general this can be represented with a prior $P_0(Z_0)$ and it is completely arbitrary, reflecting Bob's beliefs. This is what it is typically done in standard works to study `equilibration' of quantum and classical systems (with the limiting case of a quench when the prior is concentrated on one $Z_0$) \cite{Reimann2008,Goldstein2010,DAlessio2016,Mori2018}. There is no unique prior in general and so predictions, both stationary and non-stationary are generically observer dependent.

	How can Bob select the prior making use of the information he has? In the present case, Bob knows that there is a decomposition in ergodic components because he knows $\Phi$ and he would like to use this information at its best. If Bob distinguishes each state of the assumed mathematical model, a fair assumption could be that all states $Z_0$ are equally likely. But then, since he knows that each ergodic component $\Omega_\alpha$ is invariant, he assigns to each component a probability proportional to its volume as
	\begin{equation}\label{eq:bob_nu}
		\nu(\Omega_\alpha) = \frac{|\Omega_\alpha|}{\sum_\alpha |\Omega_\alpha|}\,.
	\end{equation}
	This is fine in a bounded system. Bob's intuition is that the larger the component the most probable is for $Z_0$ to be drawn from there when Alice prepares the system. It is clear that in an adversarial setting Alice might be as perverse as she likes and, in an adversarial situation, she may deliberately select an initial condition for which Bob's errors are arbitrarily large but Eq.~\eqref{eq:bob_nu} is the assignment that minimizes the future surprisal, i.e., a maximal entropy assignment \cite{Jaynes1957a,Jaynes1957b,Jaynes1968}. Obviously, Bob is free to bet anything he likes.
	
	With this choice, Bob's prediction for $Z(t^*)$ for arbitrary $t^*$ in the future, under the prior information about the knowledge of the dynamical law producing $Z(t) = \Phi_t(Z_0)$, is the limit in Eq.~\eqref{eq:bob_stat} averaged over the components, which now becomes
	\begin{equation}
		\label{eq:erg_dec}
		\mu_{\rm st} \equiv \sum_\alpha \nu(\Omega_\alpha) \mu_\alpha
	\end{equation}
	where we recall that: i) $\alpha$ labels the different ergodic components ii) $\nu(\Omega_\alpha)$ is the weight given by Bob to component $\alpha$ as in Eq.~\eqref{eq:bob_nu} iii) $\mu_\alpha$ is the $T\to \infty$ limit of the r.h.s. in Eq.~\eqref{eq:stat_pred} when $Z_0\in \Omega_\alpha$ and it is always stationary for $\mu(\alpha)$-almost all $Z_0$ \cite{Birkhoff1931,vonNeumann1932}. 
	
	Clearly, the histograms predicted with Eq.~\eqref{eq:erg_dec} have larger spreads than those predicted using Eq.~\eqref{eq:bob_choice2}. This propagates to errors made on single point estimates like average values or fluctuations of observables.
	
	As a final comment we notice that in this case of multiple ergodic components and unknown $Z_0$, a specific $Z_0$ might be `atypical' w.r.t. the prior that Bob has decided to assume: an example being the annealed mixture as in Eq.~\eqref{eq:erg_dec} derived assuming all states as equally probable and $Z_0$ lying on a manifold of dimension smaller than the available phase space. Other priors clearly lead to different typicality statements \cite{Goldstein_2012,Cerino_2016,  BALDOVIN20251}. Hence, any typicality statement seems to be bound to the choice of these priors. 
	On the other hand, in the case $Z_0$ is perfectly known, typicality of $Z_0$ is out of question as the measure on the r.h.s. of Eq.~\eqref{eq:stat_pred} is supported on the orbit $\Phi_t(Z_0)$. 
	
	\section{Single particle learning}
	Alice prepares a single particle moving on a ring of radius $R$ with conserved energy $E = \frac{1}{2}m\omega_0^2 R^2$.  We assume no force is present so that $\omega_0$ is constant in time. The motion is periodic with period $\tau = 2\pi/\omega_0$. 
	
	Then Bob is handed the system at some later time and, as we explained in the main text, he is uncertain about his future, see Eq.~\eqref{eq:bob_future} and Fig.~\ref{fig:shift}. For Bob, the dynamics is $\theta(t) = \omega_0 t + \theta_0$ and $\omega(t)=\omega_0$. Carrying out the time integral in Eq.~\eqref{eq:stat_pred} for the state $Z(t) = (\theta(t), \omega(t))$ and taking the limit $T\to \infty$, Bob finds that the stationary prediction $\mu_{\rm st}$ has a density $\rho_{\rm st}(\theta, \omega)= (2\pi)^{-1} \delta(\omega-\omega_0) \mathbb{1}(\theta\in[0,2\pi])$ because the velocity is conserved. On the other hand, the density of the microcanonical Ansatz would be $\rho_{\rm mc}(\theta, \omega)= \frac{1}{4\pi}\sum_{\sigma =\pm}\delta(\omega-\omega_\sigma)\mathbb{1}(\theta\in[0,2\pi])$ where $\omega_{\pm} = \pm \omega_0 = \pm \sqrt{2E/(m R^2)}$ (corresponding to Eq.~\eqref{eq:erg_dec}). These simple calculations are shown in \cite{supp}. These two distributions, $\rho_{\rm st}$ and $\rho_{\rm mc}$ describe two states of Bob's knowledge: the former applies when Bob knows $(\theta_0,\omega_0)$ exactly; the latter applies when he knows only $(\theta_0,E)$ is known which does not allow to reconstruct the sign of $\omega_0$ and Bob's best prediction is to average $\rho_{\rm st}$ over these two possibilities (see \cite{supp}). Neither is wrong or correct, they just describe two different states of knowledge.
	Furthermore, the microcanonical prediction $\rho_{\rm mc}$ and $\rho_{\rm st}$ give indistinguishable results for observables of the form $f(\theta, |\omega|)$ a quite large class. 
	
	As explained in the main text, Fig.~\ref{fig:gen_error} shows the optimal generalization error Eq.~\eqref{eq:kl_dec} as a function of the prediction horizon $T$ when Bob's task is to find the distribution of the full signal $Z(t) = (\theta(t), \omega(t))$ in two cases: i) when the IC $(\theta_0,\omega_0)$ is known and ii) when only $(\theta_0,E)$ is known. See Eq.~\eqref{eq:ent_main} in the main text. 
	In case ii), Bob ignores ${\rm sign}(\omega_0)$ and arrives at a larger generalization error at large $T$ (coinciding with the value of the Boltzmann entropy based on $\rho_{\rm mc}$). The information gain is given quantitatively by the KL divergence as $\Delta S \equiv D_{\rm KL}(\rho_{\rm st}||\rho_{\rm mc})=1$ bit. 
	
	Finally, microscopic oscillations in the generalization error in Fig.~\ref{fig:gen_error} stemming from Eq.~\eqref{eq:ent_main} make the entropy rate change sign and are similar to those found in \cite{Safranek2019,Safranek2021}. They can be interpreted from a learning perspective: when Bob observes samples calculated from $Z(t) = \Phi_t(Z_0)$ at exactly $T/\tau=1$ he predicts the uniform density for the distribution of $\theta$ because, by sampling, he finds the system spending equal time at all angular intervals. But during the second lap, i.e., for $T/\tau < t < 2T/\tau$, the particle will take time $\tau$ to explore the full circle again. Thus, constructing the time average as in Eq.~\eqref{eq:stat_pred} at each lap momentarily deviates from the uniform prediction reducing the learned information and causing the asymmetric oscillating dips observed in Eq.~\ref{fig:gen_error}. See \cite{supp} for details.
	
	Now, in a system of $N$ uniformly rotating particles with \emph{sufficiently spread} individual initial conditions, for the prediction of the distribution of the \emph{global state} $Z$, what matters is the Poincar\'e recurrence time $\tau\equiv\tau_{\rm Poi} \sim e^{cN}$ \cite{poincare1890, HOLMES1990137, BarrowGreen1997, barreira2006}: the plateau of the error $L^*_T$ needs exponential time to be reached meaning learning $N$ different degrees of freedom takes an exponentially large time by sampling. Notice that $\tau_{\rm Poi}$ depends on the IC, a fact often neglected. On the other hand, learning the distribution or the expectation value of a linear statistics $N^{-1}\sum_{i=1}^N f(z_i)$ where $z_i$ are the elementary degrees of freedom of a system of $N$ identical particles is much easier: the linear statistics is invariant under permutations and samples particles in space uniformly at random further reducing the error and the equilibration time. Intuitively this is because it's enough that only one out of the $N$ particles recurs at a given time to the initial state of one of the other particles. Nevertheless, the issue requires care and stationary distribution still depends on the IC \cite{DelVecchio_future}.
	
	\section{Entropy and mutual information}
	In the main text we stated that the optimum of log-loss in Eq.~\eqref{eq:opt_gen_err} corresponds to the mutual information $I(\Phi_{s_0}(Z_0),s_0)$ between the random variable $\Phi_{s_0}(Z_0)$ and the initial time shift $s_0$ in Fig.~\ref{fig:shift}. Here we would like to show this fact.
	
	By definition of mutual information we have $I(\Phi_{s_0}(Z_0), s_0) = S(\Phi_{s_0}(Z_0)) - S(\Phi_{s_0}(Z_0)|s_0)$ \cite{mezard_montanari2009}. The conditional entropy piece gives $ S(\Phi_{s_0}(Z_0)|s_0)=0$ because if Bob knew $s_0$ then $\Phi_{s_0}(Z_0)$ would be deterministic and perfectly known to him. As discussed in the text, sampling at i.i.d. times $\{t_i\}_{i=1}^M$ is equivalent to drawing $s_0$ in $[0,T]$ uniformly at random. Hence the mutual information between the hidden time origin and the present value of the signal (from the point of view of Bob) simplifies to $S[\rho_{T,Z_0}]$. 
	Consequently, the learning curve quantifies how much we learn about the hidden past of the signal as the prediction horizon $T$ grows.
	
	Of course, information quantities for continuous distributions are sometimes ill-defined. This limit is unphysical and one should always use probabilities of bins as in Eq.~\eqref{eq:stat_pred} with $A=\dd z$. In this sense, one should interpret the derivations above. Indeed, recently a coarse graining approach to entropy was introduced to cope with this problem \cite{Safranek2019, Safranek2021}.
	To appreciate the point, assume that a density exists $\mu(\dd z)=\rho(z)\dd z$. The Shannon entropy estimated from sampling in Eq.~\eqref{eq:emp_freq} is $\approx -\sum_{i} \rho(z_i)\dd z  \log(\rho(z_i) \dd z)$ where $z_i$ is any point in a bin of size $|\dd z|$. This is clearly only defined up to a constant shift in the entropy $\propto \log( \dd z)$. Yet taking the $D_{\rm KL}$ \cite{KullbackLeibler1951} as loss function in Eq.~\eqref{eq:kl_dec} resolves the problem as the shift disappears: $D_{\rm KL}(\rho || \sigma) \approx \sum_{i}\rho(z_i)\dd z \log(\rho(z_i) / \sigma(z_i))$ which is well defined as $\dd z \to 0$. This is the well known statement that only entropy differences have meaning. As a final comment on the choice of coordinates $z$ on which the entropy depends criticized in \cite{BALDOVIN20251}, we notice that the coordinates are selected by the particular measurement apparatus.

\end{document}


\title{Supplementary Material for ``On the relaxation problem in statistical mechanics''}
	\author{Giuseppe Del Vecchio Del Vecchio}
	\affiliation{Laboratoire de Physique de l'Ecole Normale Superieure, CNRS, ENS, PSL University, Paris, France}
	\maketitle
	\onecolumngrid
	
	In this Supplementary Material we give the calculations supporting the results quoted in the main text. In Sec.~\ref{sec:known_ic} we consider the ring when Bob knows the exact IC $Z_0=(\theta_0,\omega_0)$ and derive the finite-$T$ joint probability density of the full signal $Z(t)=(\theta(t),\omega(t))$, its angular marginal, and its stationary $T\to\infty$ limit. In Sec.~\ref{sec:unknown_ic} we consider incomplete knowledge of the IC and show, in particular, how Bob's prediction changes when only the conserved energy $E$ is communicated and how the microcanonical law is recovered. In Sec.~\ref{sec:logloss} we introduce finite measurement resolution and compute the log-loss generalization error, its entropy representation, the finite-$T$ learning curve, and its large-$T$ behavior. In Sec.~\ref{sec:global_relaxation} we show directly that convergence of the full recorded probability law implies convergence of its marginals, all bounded recorded expectations, and its finite-resolution entropy. Finally, in Sec.~\ref{sec:time_reversal} we study forward and backward sampling for a general time-reversal invariant dynamics, explain the role of the measurement bins, and discuss both the ring and special ICs for which the two time directions can have different limiting occupation measures. Throughout, we keep the observation horizon $T$ finite and take $T\to\infty$ only after the finite-$T$ prediction has been obtained.
	
	\section{Known initial condition}\label{sec:known_ic}
	The finite $T$ inferred joint p.d.f. is given by
	\begin{equation}\label{eq:rhoT_ring1}
		\rho_T(\theta, \omega) = \delta(\omega - \omega_0) \frac{1}{2\pi}\begin{cases}
			\frac{n+1}{n+r} & \theta \in I_{r}(\theta_0)\\
			\frac{n}{n+r} & \theta \notin I_{r}(\theta_0)
		\end{cases}
	\end{equation}
	where $n = \lfloor\frac{T}{\tau}\rfloor$ is the integer part, $r=\{\frac{T}{\tau}\}$ is the decimal part, $\tau = \frac{2\pi}{\omega_0}$ is the period and $I_r(\theta_0) = \{2\pi u + \theta_0 \mod 2\pi: 0\le u < r\}$ is the arc traversed in one incomplete revolution. Notice how this depends on the IC $\theta_0$ and $\omega_0$. From the joint p.d.f. in Eq.~\eqref{eq:rhoT_ring1} we can compute everything else. The calculation proceeds as follows. 
	
	The dynamical rule $\Phi$ that Alice communicates to Bob evolves the IC as $\Phi_t(\theta_0,\omega_0)=(\theta(t), \omega(t))$ where
	\begin{equation}\label{eq:dynamics_ring}
		\theta(t) = \theta_0 + \omega_0  t\mod 2\pi\quad \text{and} \quad \omega(t) = \omega_0\,.
	\end{equation}
	To calculate the p.d.f. $\rho_T(\theta, \omega)$ we differentiate the occupation time on right hand side (r.h.s.) of Eq.~(5) of the main text w.r.t. $\theta$ and $\omega$. This gives the local time \cite{Levy1940, majumdar2006}
	\begin{equation}
		\label{eq:local_time}
		\rho_T(\theta, \omega) = \frac{1}{T}\int_0^T \delta(\theta-\theta(t)) \delta(\omega-\omega(t))\dd t =  \frac{1}{T}\delta(\omega-\omega_0)\int_0^T \delta(\theta-\omega_0 t - \theta_0\mod 2\pi)\dd t
	\end{equation}
	The first equality in Eq.~\eqref{eq:local_time} is the density form of the occupation measure in Eq.~(5) of the main text. To obtain the second equality we substitute the deterministic dynamics in Eq.~\eqref{eq:dynamics_ring}: since $\omega(t)=\omega_0$, the factor $\delta(\omega-\omega(t))$ becomes $\delta(\omega-\omega_0)$ and can be taken outside the time integral, while $\theta(t)=\theta_0+\omega_0t\mod 2\pi$ gives the remaining delta function.
	Notice that the local time using delta functions as in Eq.~\eqref{eq:local_time} is well defined only in one dimension, otherwise one either needs to compute the occupation time or needs a regularization \cite{majumdar2006, Del_vecchio_del_vecchio_2025}.
	
	Now, since every real number $x$ can be written as its integer part plus its fractional part $x = \lfloor x \rfloor + \{x\}$ where, we can write the length of the observation window as
	\begin{equation}\label{eq:split}
		T = (n + r)\tau\quad \text{where} \quad n = \left\lfloor\frac{T}{\tau}\right\rfloor \text{and} \quad r=\left\{\frac{T}{\tau}\right\}
	\end{equation}
	as already defined below Eq.~\eqref{eq:rhoT_ring1}. Changing variables to $u = t / \tau$ in Eq.~\eqref{eq:local_time} we write
	\begin{equation}\label{eq:rhoT_2}
		\rho_T(\theta, \omega)=\frac{1}{n+r}\delta(\omega-\omega_0)\int_0^{n+r} \delta\left(\theta-\theta_0-2\pi u\mod 2\pi\right)\dd u
	\end{equation}
	where we have used Eq.~\eqref{eq:split} to express the denominator $T$ in terms of $n$ and $r$.
	Splitting the integral in Eq.~\eqref{eq:rhoT_2} we obtain
	\begin{align}
		\label{eq:int_split}
		\int_0^{n+r}\delta\left(\theta-\theta_0-2\pi u\mod 2\pi\right)\dd u
		&= \int_0^{n}\delta\left(\theta-\theta_0-2\pi u\mod 2\pi\right)\dd u
		+\int_n^{n+r}\delta\left(\theta-\theta_0-2\pi u\mod 2\pi\right)\dd u
		\nonumber\\
		&= \int_0^{n}\delta\left(\theta-\theta_0-2\pi u\mod 2\pi\right)\dd u
		+\int_0^{r}\delta\left(\theta-\theta_0-2\pi u\mod 2\pi\right)\dd u
		\nonumber\\
		&= \frac{n}{2\pi}
		+\int_0^{r}\delta\left(\theta-\theta_0-2\pi u\mod 2\pi\right)\dd u .
	\end{align}
	The first equality in Eq.~\eqref{eq:int_split} simply divides the integration interval $[0,n+r]$ into the complete part $[0,n]$ and the remaining part $[n,n+r]$. In the second equality we shift the variable by the integer $n$ in the second integral. The integrand is periodic in $u$ with period $1$, so this turns the interval $[n,n+r]$ into $[0,r]$ without changing the integrand. In the third equality we use that the first integral contains exactly $n$ complete periods, each contributing $1/(2\pi)$.
	To recover Eq.~\eqref{eq:rhoT_ring1}, the remaining integral in the last line of Eq.~\eqref{eq:int_split} gives $1/(2\pi)$ if $\theta \in I_r(\theta_0)=\{2\pi u + \theta_0: 0\le u < r\}$ and it is otherwise $0$. Hence, substituting Eq.~\eqref{eq:int_split} into Eq.~\eqref{eq:rhoT_2} gives Eq.~\eqref{eq:rhoT_ring1}. 
	
\begin{figure}[h]
		\centering
		\includegraphics[width=0.95\linewidth]{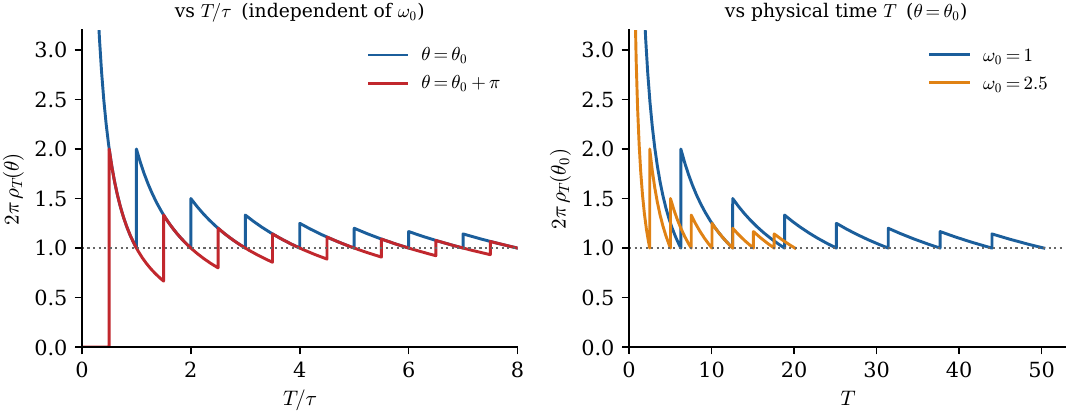}
		\caption{Finite-horizon angle density $\rho_T(\theta)$ in Eq.~\eqref{eq:rhoT_theta}. We plot $2\pi\,\rho_T(\theta)=\big(n+\mathbb{1}(\theta\in I_r(\theta_0))\big)/(n+r)$, which relaxes to the stationary value $1$ in Eq.~\eqref{eq:rho_st_theta} (dotted line). \emph{Left}: as a function of $T/\tau$ for two observation angles, $\theta=\theta_0$ (always inside the freshly swept arc $I_r(\theta_0)$, upper branch) and $\theta=\theta_0+\pi$; in these units the curve is independent of $\omega_0$, all the dependence entering through $\tau=2\pi/\omega_0$ defined in Eq.~\eqref{eq:split}. \emph{Right}: the same quantity at $\theta=\theta_0$ versus the physical horizon $T$ for two angular velocities $\omega_0=1,\,2.5$; a larger $\omega_0$ (shorter period $\tau$) relaxes faster in real time. The sawtooth reflects each lap momentarily over- or under-visiting a given angle, in agreement with the oscillations discussed around Fig.~2 of the main text. The density diverges as $T/\tau\to0$ (short horizon, sharply peaked occupation), so the vertical axis is capped for readability.}
		\label{fig:jpdfto}
	\end{figure}

	The joint law $\rho_T(\theta,\omega)$ in Eq.~\eqref{eq:rhoT_ring1} is singular w.r.t. $\omega$ because the continuous velocity is exactly conserved, as stated in Eq.~\eqref{eq:dynamics_ring} (and as occurs in integrable models \cite{Essler_2016}). Integrating out $\omega$ in Eq.~\eqref{eq:rhoT_ring1} gives the angular density
	\begin{equation}\label{eq:rhoT_theta}
		\rho_T(\theta) \equiv \int_{\mathbb{R}} \rho_T(\theta, \omega)\dd \omega = \frac{1}{2\pi}\begin{cases}
			\frac{n+1}{n+r} & \theta \in I_{r}(\theta_0)\\
			\frac{n}{n+r} & \theta \notin I_{r}(\theta_0)
		\end{cases}
	\end{equation}
	In Eq.~\eqref{eq:rhoT_theta}, the first equality defines the angular marginal by integrating the joint law in Eq.~\eqref{eq:rhoT_ring1} over $\omega$. The second equality follows because the integral of $\delta(\omega-\omega_0)$ over $\mathbb{R}$ is one. The resulting density still depends on $\omega_0$ through $n$ and $r$, defined from $T/\tau$ in Eq.~\eqref{eq:split}. As $T\to \infty$, $n \to \infty$ and we obtain Bob's stationary prediction
	\begin{equation}\label{eq:rho_st_theta}
		\rho_{\rm st}(\theta) = \lim_{T\to \infty}\rho_T(\theta) = \frac{1}{2\pi}
	\end{equation}
	The first equality in Eq.~\eqref{eq:rho_st_theta} defines the stationary angular density as the $T\to\infty$ limit of Eq.~\eqref{eq:rhoT_theta}. In this limit both factors $(n+1)/(n+r)$ and $n/(n+r)$ tend to one, which gives the second equality $\rho_{\rm st}(\theta)=1/(2\pi)$. Thus the stationary angular distribution is uniform. 
	A plot of the angular density $\rho_T(\theta)$ defined in Eq.~\eqref{eq:rhoT_theta} is provided in Fig.~\ref{fig:jpdfto}.

	\section{Unknown initial condition}\label{sec:unknown_ic}
	In Sec.~\ref{sec:known_ic} Alice communicated to Bob both the rule $\Phi$ and the exact IC $Z_0=(\theta_0,\omega_0)$. We now treat the physically more common situation, anticipated in the main text, in which Bob is told the rule and the conserved \emph{energy} $E=\frac12 m\omega_0^2R^2$ but \emph{not} the IC itself. Knowing $E$ fixes the speed $|\omega_0|=\sqrt{2E/(mR^2)}$, hence the period $\tau=2\pi/|\omega_0|$ used in Eq.~\eqref{eq:split}, but leaves two things undetermined: the initial angle $\theta_0$ and the \emph{sign} of $\omega_0$, i.e.\ the sense of rotation. As explained around Eq.~(12) of the main text, Bob must now assign a prior over these missing data and average the finite-$T$ prediction in Eq.~\eqref{eq:rhoT_ring1} accordingly.
	
	Being maximally noncommittal [Eq.~(11) of the main text], Bob takes $\theta_0$ uniform on $[0,2\pi)$ and the two rotation senses equally likely. The energy shell is the union of two ergodic components $\Omega_\pm=\{(\theta,\pm|\omega_0|)\}$, each an invariant circle; by the reflection symmetry $\omega\to-\omega$ they have equal volume, so $\nu(\Omega_+)=\nu(\Omega_-)=\tfrac12$ in Eq.~(11) of the main text. Bob's prediction is therefore
	\begin{equation}\label{eq:rhoT_unk}
		\rho_T^{\rm unk}(\theta,\omega)=\sum_{\sigma=\pm}\tfrac12\int_0^{2\pi}\frac{\dd\theta_0}{2\pi}\,\rho_T^{(\sigma)}(\theta,\omega\,|\,\theta_0)\,,
	\end{equation}
	where $\rho_T^{(\sigma)}$ is the known-IC law in Eq.~\eqref{eq:rhoT_ring1} with $\omega_0$ replaced by $\sigma|\omega_0|$. Two independent simplifications occur.
	
	\emph{(i) Unknown $\theta_0$ erases the transient.} Fix the sign and integrate the angular density in Eq.~\eqref{eq:rhoT_theta} over $\theta_0$. Since $I_r(\theta_0)$, defined below Eq.~\eqref{eq:rhoT_ring1}, is an arc of length $2\pi r$ whose position is set by $\theta_0$, the probability that a uniformly placed arc covers a fixed $\theta$ is exactly $r$, i.e.\ $\int_0^{2\pi}\frac{\dd\theta_0}{2\pi}\,\mathbb{1}(\theta\in I_r(\theta_0))=r$. Hence
	\begin{align}\label{eq:theta0_avg}
		\int_0^{2\pi}\frac{\dd\theta_0}{2\pi}\,\rho_T(\theta\,|\,\theta_0)
		&=\frac{1}{2\pi}\left[\frac{n+1}{n+r}\,r+\frac{n}{n+r}(1-r)\right]\nonumber\\
		&=\frac{1}{2\pi}\,.
	\end{align}
	The first equality in Eq.~\eqref{eq:theta0_avg} follows from Eq.~\eqref{eq:rhoT_theta}: for fixed $\theta$, the fraction of values of $\theta_0$ for which $\theta\in I_r(\theta_0)$ is $r$, while the complementary fraction is $1-r$. In the second equality the numerator simplifies as $r(n+1)+(1-r)n=n+r$, which cancels the denominator $n+r$. Therefore the result is $1/(2\pi)$ \emph{at every finite $T$}. Not knowing where the particle started, Bob predicts the uniform angular law immediately: there is nothing left to learn about $\theta$ and the relaxation described in Sec.~\ref{sec:known_ic} disappears.
	
	\emph{(ii) Unknown sign is a static bit.} Because $\omega(t)=\omega_0$ is conserved by Eq.~\eqref{eq:dynamics_ring}, the velocity marginal
	\begin{equation}\label{eq:vel_marg}
		\rho^{\rm unk}(\omega)=\tfrac12\,\delta(\omega-|\omega_0|)+\tfrac12\,\delta(\omega+|\omega_0|)
	\end{equation}
	is independent of $T$: the actual sign of the system in the box is never revealed by sampling from the dynamical rule $\Phi_t$ and the associated uncertainty is a rigid one bit.
	
	Combining (i)--(ii), Bob's prediction in Eq.~\eqref{eq:rhoT_unk} becomes the \emph{microcanonical} law quoted in the EM,
	\begin{equation}\label{eq:rho_micro_sm}
		\rho_T^{\rm unk}(\theta, \omega)=\rho_{\rm mc}(\theta,\omega)=\frac{1}{4\pi}\sum_{\sigma=\pm}\delta(\omega-\sigma|\omega_0|)\,\mathbb{1}(\theta\in[0,2\pi))\,,
	\end{equation}
	The first equality in Eq.~\eqref{eq:rho_micro_sm} states that, after averaging over the unknown $\theta_0$ and the unknown sign in Eq.~\eqref{eq:rhoT_unk}, Bob's finite-$T$ prediction is already stationary. The second equality identifies this stationary mixture with the microcanonical law: Eq.~\eqref{eq:theta0_avg} gives the uniform factor $1/(2\pi)$ in $\theta$, while Eq.~\eqref{eq:vel_marg} gives equal weights $1/2$ to the two allowed signs of $\omega$.
	This is to be compared with $\rho_{\rm st}(\theta,\omega)=(2\pi)^{-1}\delta(\omega-\omega_0)\mathbb{1}(\theta\in[0,2\pi))$, obtained by taking $T\to\infty$ in Eq.~\eqref{eq:rhoT_ring1}. The two differ only by the sign information, $D_{\rm KL}(\rho_{\rm st}\|\rho_{\rm mc})=\log 2$, one bit, exactly the plateau gap of Fig.~2 of the main text.
	
\begin{figure}[h]
		\centering
		\includegraphics[width=0.95\linewidth]{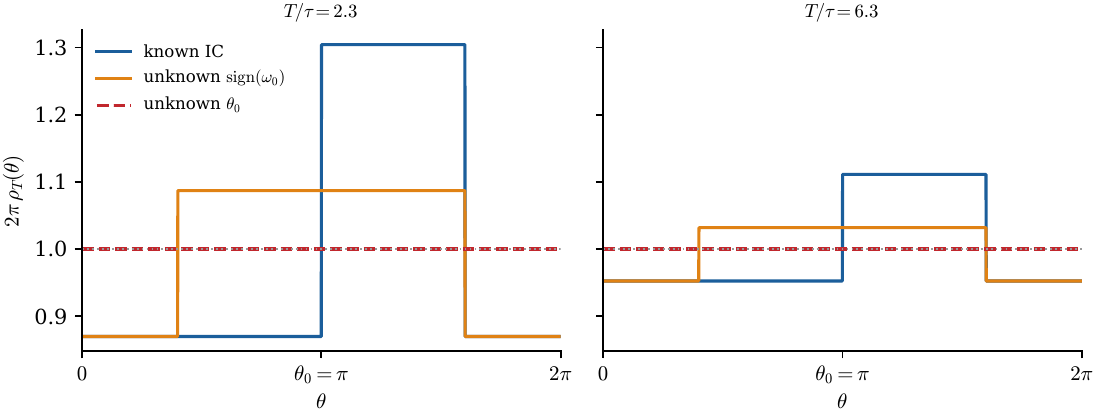}
		\caption{Angular density $2\pi\,\rho_T(\theta)$ in Eq.~\eqref{eq:rhoT_theta} under three states of Bob's knowledge, at two horizons $T/\tau=2.3$ (left) and $T/\tau=6.3$ (right), with $\theta_0=\pi$. Blue: known IC. Orange: only $\mathrm{sign}(\omega_0)$ unknown, Eq.~\eqref{eq:rho_unksign}. Dashed red: $\theta_0$ unknown, Eq.~\eqref{eq:theta0_avg}. As $T/\tau$ grows, the first two densities approach the stationary density in Eq.~\eqref{eq:rho_st_theta}.}
		\label{fig:rhoT_unknown}
	\end{figure}

	It is instructive to keep $\theta_0$ known but the sign unknown, the case underlying Fig.~2 of the main text. Then only the $\sigma$-average survives in Eq.~\eqref{eq:rhoT_unk}, and the two senses sweep the forward arc $I_r(\theta_0)$ and the backward arc $I_r^-(\theta_0)=\{\theta_0-2\pi u\bmod 2\pi:0\le u<r\}$. For $r<\tfrac12$ these do not overlap and
	\begin{equation}\label{eq:rho_unksign}
		2\pi\,\rho_T^{\rm sign}(\theta)=
		\begin{cases}
			\dfrac{2n+1}{2(n+r)}, & \theta\in I_r(\theta_0)\cup I_r^-(\theta_0)\\[2mm]
			\dfrac{n}{n+r}, & \text{otherwise}\,,
		\end{cases}
	\end{equation}
	a symmetric double step of half the excess height. Fig.~\ref{fig:rhoT_unknown} compares the known-IC density in Eq.~\eqref{eq:rhoT_theta}, the unknown-sign density in Eq.~\eqref{eq:rho_unksign}, and the uniform density obtained by averaging $\theta_0$ in Eq.~\eqref{eq:theta0_avg}. As $T/\tau$ grows, Eqs.~\eqref{eq:rhoT_theta} and \eqref{eq:rho_unksign} approach the stationary density in Eq.~\eqref{eq:rho_st_theta}. The lesson is the one anticipated in the main text: the stationary law is not a property of the ring but of Bob's information; more ignorance means a flatter, higher-entropy prediction.

	\section{Log-loss error and the learning curve}\label{sec:logloss}
	Finally we compute the generalization error for the log-loss. As explained in the EM, information is defined for the discrete outcomes recorded by a measurement apparatus. The full inferred law is the joint law $\rho_T(\theta,\omega)$ in Eq.~\eqref{eq:rhoT_ring1}, and its angular marginal $\rho_T(\theta)$ is defined in Eq.~\eqref{eq:rhoT_theta}.
	
	To regularize both continuous variables, divide the $(\theta,\omega)$ plane into bins $B_j\times C_k$, where the angular bins $B_j$ have width $\Delta\theta=2\pi/K$ and the velocity bins $C_k$ have width $\Delta\omega$. The probability of the recorded joint outcome $(j,k)$ is
	\begin{equation}\label{eq:joint_cell_probability}
		P_{jk}(T)=\int_{B_j}\dd\theta\int_{C_k}\dd\omega\,\rho_T(\theta,\omega).
	\end{equation}
	Let $C_{k_0}$ be the velocity bin containing the known value $\omega_0$. Using the joint law in Eq.~\eqref{eq:rhoT_ring1} and its angular marginal in Eq.~\eqref{eq:rhoT_theta}, Eq.~\eqref{eq:joint_cell_probability} becomes
	\begin{equation}\label{eq:joint_cell_factorization}
		P_{jk}(T)=p_j(T)\,\mathbb{1}(k=k_0),
		\qquad
		p_j(T)=\int_{B_j}\rho_T(\theta)\dd\theta.
	\end{equation}
	The first relation in Eq.~\eqref{eq:joint_cell_factorization} follows because the factor $\delta(\omega-\omega_0)$ in Eq.~\eqref{eq:rhoT_ring1} puts all the probability in the single velocity bin $C_{k_0}$. The second relation defines $p_j(T)$ as the probability of the angular bin $B_j$, obtained by integrating the angular density $\rho_T(\theta)$ in Eq.~\eqref{eq:rhoT_theta} over that bin.
	The finite-resolution entropy of the global recorded state is therefore
	\begin{align}\label{eq:global_discrete_entropy}
		S_{\Delta\theta,\Delta\omega}[\rho_T]
		&=-\sum_{j,k}P_{jk}(T)\log P_{jk}(T)\nonumber\\
		&=-\sum_jp_j(T)\log p_j(T).
	\end{align}
	In the first equality of Eq.~\eqref{eq:global_discrete_entropy} we use the definition of the Shannon entropy of the joint binned distribution $P_{jk}(T)$ introduced in Eq.~\eqref{eq:joint_cell_probability}. In the second equality we use Eq.~\eqref{eq:joint_cell_factorization}: all terms with $k\neq k_0$ vanish, while the only nonzero term for each $j$ is $P_{jk_0}(T)=p_j(T)$. Thus $\omega$ has been included in the global entropy. Since the conserved known velocity always occupies the single bin $C_{k_0}$ in Eq.~\eqref{eq:joint_cell_factorization}, its probability is one and its entropy contribution is $-1\log1=0$. Equation~\eqref{eq:global_discrete_entropy} consequently holds for every $\Delta\omega$ for which $\omega_0$ is assigned to one bin, and taking $\Delta\omega\to0$ adds no divergent term.
	
	It remains to remove the angular resolution. Equation~\eqref{eq:rhoT_theta} shows that $\rho_T(\theta)$ is constant in every angular bin that does not contain an endpoint of the arc $I_r(\theta_0)$ defined below Eq.~\eqref{eq:rhoT_ring1}. For such a bin, Eq.~\eqref{eq:joint_cell_factorization} gives $p_j(T)=\rho_T(\theta_j)\Delta\theta$ for any $\theta_j\in B_j$. Each of the two endpoint bins has probability $O(\Delta\theta)$ and contributes $O(\Delta\theta|\log\Delta\theta|)\to0$. Substitution in Eq.~\eqref{eq:global_discrete_entropy} gives
	\begin{align}
		S_{\Delta\theta,\Delta\omega}[\rho_T]
		&=-\sum_j\rho_T(\theta_j)\Delta\theta
		\log\!\left[\rho_T(\theta_j)\Delta\theta\right]+o(1)\nonumber\\
		&=-\int_0^{2\pi}\rho_T(\theta)\log\rho_T(\theta)\dd\theta
		-\log\Delta\theta+o(1),
		\label{eq:entropy_regularized}
	\end{align}
	In the first equality of Eq.~\eqref{eq:entropy_regularized} we substitute $p_j(T)=\rho_T(\theta_j)\Delta\theta$ into Eq.~\eqref{eq:global_discrete_entropy}; the two bins containing the endpoints of $I_r(\theta_0)$ contribute only to the $o(1)$ term. To obtain the second equality we expand $\log[\rho_T(\theta_j)\Delta\theta]=\log\rho_T(\theta_j)+\log\Delta\theta$. The sum containing $\log\rho_T(\theta_j)$ becomes the integral as $\Delta\theta\to0$, while the term proportional to $\log\Delta\theta$ gives $-\log\Delta\theta$ because $\sum_j\rho_T(\theta_j)\Delta\theta\to1$. Here $o(1)$ denotes terms that vanish as $\Delta\theta\to0$. Thus the quantity plotted in Fig.~2 of the main text is the finite part of the global entropy,
	\begin{equation}
		S[\rho_T]\equiv\lim_{\substack{\Delta\theta\to0\\\Delta\omega\to0}}
		\left(S_{\Delta\theta,\Delta\omega}[\rho_T]+\log\Delta\theta\right)
		=-\int_0^{2\pi}\rho_T(\theta)\log\rho_T(\theta)\dd\theta.
		\label{eq:entropy_continuous}
	\end{equation}
	The first equality in Eq.~\eqref{eq:entropy_continuous} defines $S[\rho_T]$ by adding $\log\Delta\theta$ to the finite-resolution entropy, thereby removing the term $-\log\Delta\theta$ identified in Eq.~\eqref{eq:entropy_regularized}. The second equality follows by substituting Eq.~\eqref{eq:entropy_regularized}: the two $\log\Delta\theta$ terms cancel and the $o(1)$ term vanishes as $\Delta\theta\to0$.
	
	We now evaluate Eq.~\eqref{eq:entropy_continuous} in closed form. Equation~\eqref{eq:rhoT_theta} gives a constant angular density on the arc $I_r(\theta_0)$ and another constant outside it. With $n,r$ defined in Eq.~\eqref{eq:split}, write these two factors as
	\begin{equation}\label{eq:ab_definition}
		a\equiv\frac{n+1}{n+r},
		\qquad
		b\equiv\frac{n}{n+r}.
	\end{equation}
	The corresponding densities are $a/(2\pi)$ on the arc, whose length is $2\pi r$ by the definition below Eq.~\eqref{eq:rhoT_ring1}, and $b/(2\pi)$ on its complement. Consequently,
	\begin{align}\label{eq:entropy_ring}
		L_T^*=S[\rho_T]
		&=-r\,a\log\frac{a}{2\pi}-(1-r)\,b\log\frac{b}{2\pi}\nonumber\\
		&=\log(2\pi)-r\,a\log a-(1-r)\,b\log b\nonumber\\
		&=\log(2\pi)-r\,\frac{n+1}{n+r}\log\frac{n+1}{n+r}-(1-r)\,\frac{n}{n+r}\log\frac{n}{n+r}\,.
	\end{align}
	where the normalization of the two pieces is
	\begin{equation}\label{eq:ab_normalization}
		r\,a+(1-r)\,b
		=\frac{r(n+1)+(1-r)n}{n+r}
		=\frac{n+r}{n+r}=1.
	\end{equation}
	In the first equality of Eq.~\eqref{eq:ab_normalization} we add the probability $ra$ carried by the arc and the probability $(1-r)b$ carried by its complement. The second equality substitutes $a$ and $b$ from Eq.~\eqref{eq:ab_definition}. The third equality uses $r(n+1)+(1-r)n=n+r$, and the last equality is the resulting normalization.
	
	We can now spell out the three steps in Eq.~\eqref{eq:entropy_ring}. The first equality evaluates the integral in Eq.~\eqref{eq:entropy_continuous} separately on the arc $I_r(\theta_0)$ and on its complement: their probabilities are $ra$ and $(1-r)b$, while their densities are $a/(2\pi)$ and $b/(2\pi)$. To obtain the second equality we expand $\log[a/(2\pi)]=\log a-\log(2\pi)$ and similarly for $b$; the two terms proportional to $\log(2\pi)$ combine to a single $\log(2\pi)$ by Eq.~\eqref{eq:ab_normalization}. The third equality follows by substituting $a$ and $b$ from Eq.~\eqref{eq:ab_definition}. We understand $0\log0$ as $0$. Equations~\eqref{eq:entropy_ring} and \eqref{eq:ab_normalization} prove Eq.~(10) of the main text for known $\omega_0$.
	
	The same global binning shows explicitly that the generalization gap has no divergent resolution-dependent constant. From the stationary angular density in Eq.~\eqref{eq:rho_st_theta}, the stationary joint-bin probabilities are
	\begin{equation}\label{eq:stationary_cell_probability}
		Q_{jk}=q_j\,\mathbb{1}(k=k_0),
		\qquad
		q_j=\int_{B_j}\rho_{\rm st}(\theta)\dd\theta
		=\frac{1}{K}=\frac{\Delta\theta}{2\pi}.
	\end{equation}
	In Eq.~\eqref{eq:stationary_cell_probability}, the first equality has the same factorized form as Eq.~\eqref{eq:joint_cell_factorization} because the known velocity remains in the single bin $C_{k_0}$. The second equality defines the stationary angular-bin probability $q_j$. The third equality uses the uniform stationary density $\rho_{\rm st}(\theta)=1/(2\pi)$ from Eq.~\eqref{eq:rho_st_theta}, so every angular bin has probability $1/K$. The last equality uses the bin width $\Delta\theta=2\pi/K$.
	
	Using $P_{jk}(T)$ from Eq.~\eqref{eq:joint_cell_factorization} and $Q_{jk}$ from Eq.~\eqref{eq:stationary_cell_probability}, the global KL divergence is
	\begin{align}\label{eq:dkl_ring}
		D_{\rm KL}(\rho_T\|\rho_{\rm st})
		&=\lim_{\substack{\Delta\theta\to0\\\Delta\omega\to0}}
		\sum_{j,k:P_{jk}>0}P_{jk}(T)\log\frac{P_{jk}(T)}{Q_{jk}}
		\nonumber
		\\
		&=\lim_{\Delta\theta\to0}\sum_j p_j(T)\log\frac{p_j(T)}{q_j}
		\nonumber
		\\
		&=\int_0^{2\pi}\rho_T(\theta)\log\!\left[2\pi\rho_T(\theta)\right]\dd\theta
		=\log(2\pi)-S[\rho_T]
		=r\,a\log a+(1-r)\,b\log b\ge0.
	\end{align}
	The first equality in Eq.~\eqref{eq:dkl_ring} is the definition of the KL divergence of the finite-resolution joint-bin probabilities, followed by the resolution limit. In the second equality we use Eqs.~\eqref{eq:joint_cell_factorization} and \eqref{eq:stationary_cell_probability}: only the velocity bin $k=k_0$ is occupied, so the sum over $k$ disappears and there is no remaining $\Delta\omega$ dependence. The third equality is the $\Delta\theta\to0$ limit of the angular sum, using $p_j(T)=\rho_T(\theta_j)\Delta\theta+o(\Delta\theta)$ and $q_j=\Delta\theta/(2\pi)$ from Eq.~\eqref{eq:stationary_cell_probability}. To obtain the fourth equality we expand $\log[2\pi\rho_T(\theta)]=\log(2\pi)+\log\rho_T(\theta)$, use $\int_0^{2\pi}\rho_T(\theta)\dd\theta=1$, and then use the definition of $S[\rho_T]$ in Eq.~\eqref{eq:entropy_continuous}. The last equality follows by substituting Eq.~\eqref{eq:entropy_ring}; the final inequality is the non-negativity of KL divergence recalled below Eq.~(8) of the main text.
	
	At every integer horizon ($r=0$), Eq.~\eqref{eq:dkl_ring} gives $D_{\rm KL}=0$. Between laps the gap is positive. To obtain its large-$n$ form, set $x=n+r$, with $n,r$ defined in Eq.~\eqref{eq:split}. Equation~\eqref{eq:ab_definition} then gives $a=1+(1-r)/x$ and $b=1-r/x$. Using $(1+u)\log(1+u)=u+u^2/2+O(u^3)$ and $(1-u)\log(1-u)=-u+u^2/2+O(u^3)$ in Eq.~\eqref{eq:dkl_ring}, the terms proportional to $x^{-1}$ cancel and
	\begin{equation}\label{eq:dkl_asymptotic}
		D_{\rm KL}
		=\frac{r(1-r)^2+(1-r)r^2}{2x^2}+O(x^{-3})
		=\frac{r(1-r)}{2(n+r)^2}+O(n^{-3}).
	\end{equation}
	In the first equality of Eq.~\eqref{eq:dkl_asymptotic} we substitute the large-$x$ expansions of $a\log a$ and $b\log b$ into the last line of Eq.~\eqref{eq:dkl_ring}; the terms of order $x^{-1}$ cancel. In the second equality we factor $r(1-r)$ in the numerator and use $(1-r)+r=1$ together with $x=n+r$. Since $0\le r<1$, $O(x^{-3})$ is also $O(n^{-3})$ for large $n$. Thus Eq.~\eqref{eq:dkl_asymptotic} shows that the dips decay with a $1/n^2$ envelope. At fixed resolution and fixed preparation, the conditional entropy at known $s_0$ is zero, and the mutual-information identity derived in the EM gives
	\begin{equation}\label{eq:mutual_information_regularized}
		I(\Phi_{s_0}(Z_0),s_0)=S_{\Delta\theta,\Delta\omega}[\rho_T]
		=S[\rho_T]-\log\Delta\theta+o(1).
	\end{equation}
	In the first equality of Eq.~\eqref{eq:mutual_information_regularized}, the mutual information equals the finite-resolution entropy because, once $s_0$ is known, the recorded bin is determined by $\Phi$ and $Z_0$ and the corresponding conditional entropy is zero. The second equality is precisely the finite-resolution relation derived in Eq.~\eqref{eq:entropy_regularized}. Thus Eq.~\eqref{eq:mutual_information_regularized} shows that Fig.~2 of the main text has the same $T$ dependence as the finite-resolution mutual information but is shifted by the constant $\log\Delta\theta$. The plotted plateau is $\log(2\pi)$ by Eq.~\eqref{eq:entropy_ring}, whereas the entropy of the $K$ occupied joint bins is $\log K=\log(2\pi)-\log\Delta\theta$.
	
	Finally, suppose $(\theta_0,E)$ is known but the sign $\sigma={\rm sign}(\omega_0)$ is not. This is the equal mixture of the two known-IC laws in Eq.~\eqref{eq:rhoT_ring1}, as obtained from Eq.~\eqref{eq:rhoT_unk} by keeping $\theta_0$ fixed. Assume that the velocity bins $C_{k_+}$ and $C_{k_-}$ containing $+|\omega_0|$ and $-|\omega_0|$ are distinct. Define
	\begin{equation}\label{eq:sign_branch_probability}
		p_j^{(\sigma)}(T)
		=\int_{B_j}\dd\theta\int_{C_{k_\sigma}}\dd\omega\,
		\rho_T^{(\sigma)}(\theta,\omega),
		\qquad
		P_{j k_\sigma}^{\rm sign}(T)=\frac{p_j^{(\sigma)}(T)}{2},
	\end{equation}
	where $\rho_T^{(\sigma)}$ is the known-IC law in Eq.~\eqref{eq:rhoT_ring1} with $\omega_0$ replaced by $\sigma|\omega_0|$. The entropy of each angular branch is
	\begin{equation}\label{eq:sign_branch_entropy}
		S_{\Delta\theta}^{(\sigma)}
		=-\sum_j p_j^{(\sigma)}(T)\log p_j^{(\sigma)}(T).
	\end{equation}
	Using the joint probabilities in Eq.~\eqref{eq:sign_branch_probability} and the branch entropies in Eq.~\eqref{eq:sign_branch_entropy}, the global discrete entropy is exactly
	\begin{align}\label{eq:unknown_sign_entropy}
		-\sum_{\sigma=\pm}\sum_j\frac{p_j^{(\sigma)}(T)}{2}
		\log\frac{p_j^{(\sigma)}(T)}{2}
		&=\log2+\frac12\left(S_{\Delta\theta}^{(+)}+S_{\Delta\theta}^{(-)}\right)\nonumber\\
		&=S[\rho_T]-\log\Delta\theta+\log2+o(1).
	\end{align}
	To obtain the first equality in Eq.~\eqref{eq:unknown_sign_entropy}, we write $\log[p_j^{(\sigma)}(T)/2]=\log p_j^{(\sigma)}(T)-\log2$. The $\log2$ part gives one factor $\log2$ because each branch is normalized, while the remaining terms are one half of the two branch entropies defined in Eq.~\eqref{eq:sign_branch_entropy}. In the second equality we use Eq.~\eqref{eq:entropy_regularized} for each branch. Reflection reverses the arc in Eq.~\eqref{eq:rhoT_theta} but leaves its continuous entropy unchanged, so the two branches have the same finite part $S[\rho_T]$. Therefore the unknown sign adds $\chi=\log2$ at every $T$, proving the second case of Eq.~(10) of the main text. The finite part of the global entropy has plateau $\log(4\pi)$, while the curve separation is $\log2$ nats, namely one bit. If $\Delta\omega$ is too large to distinguish the two signs, the probabilities in Eq.~\eqref{eq:sign_branch_probability} must instead be added within the same velocity bin, and the $\log2$ term does not follow. The orange curve in Fig.~2 of the main text assumes that the signs are resolved.
	\begin{figure}[h]
		\centering
		\includegraphics[width=0.95\linewidth]{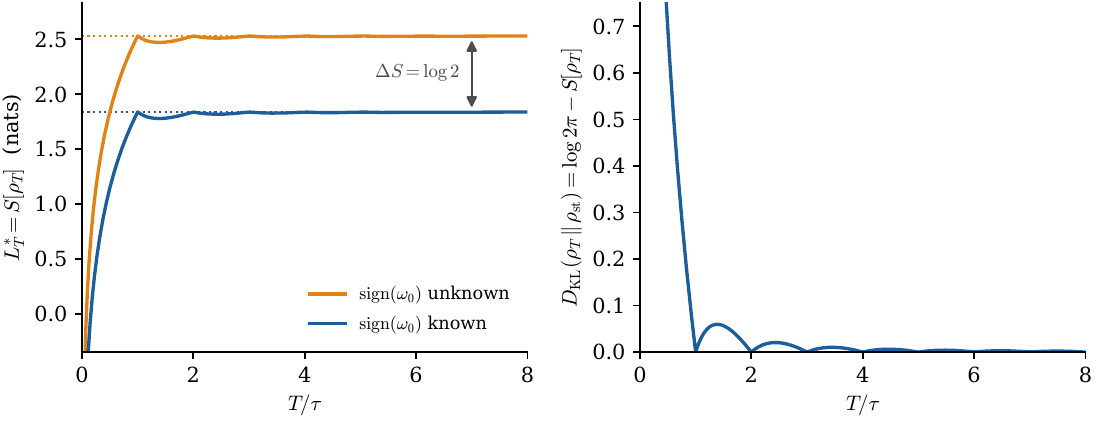}
		\caption{Log-loss generalization error for the ring. The left panel shows the finite part of the global entropy in Eq.~\eqref{eq:entropy_continuous}; Eq.~\eqref{eq:entropy_regularized} gives the finite-resolution value. The orange curve assumes that $\Delta\omega$ resolves the two velocity signs, giving Eq.~\eqref{eq:unknown_sign_entropy}. The right panel shows the global generalization gap in Eq.~\eqref{eq:dkl_ring}, whose decay is derived in Eq.~\eqref{eq:dkl_asymptotic}.}
		\label{fig:logloss}
	\end{figure}
	
	\section{Consequences of global relaxation}\label{sec:global_relaxation}
	We now spell out the elementary consequence of global relaxation quoted in the main text, keeping the same finite-resolution notation used above. The angular bins $B_j$ and the velocity bins $C_k$ are those introduced before Eq.~\eqref{eq:joint_cell_probability}. Their joint probabilities at finite $T$ are $P_{jk}(T)$, defined in Eq.~\eqref{eq:joint_cell_probability}. Let $Q_{jk}$ denote the corresponding stationary joint-bin probabilities, as in Eq.~\eqref{eq:stationary_cell_probability}. Global relaxation of the full recorded distribution means
	\begin{equation}\label{eq:full_law_convergence}
		P_{jk}(T)\longrightarrow Q_{jk}
		\qquad\text{for every recorded bin }B_j\times C_k.
	\end{equation}
	At fixed measurement resolution there are only finitely many bins. Therefore Eq.~\eqref{eq:full_law_convergence} implies
	\begin{equation}\label{eq:l1_from_full_law}
		\sum_{j,k}\left|P_{jk}(T)-Q_{jk}\right|
		\longrightarrow0.
	\end{equation}
	Indeed, every term in the finite sum tends to zero by Eq.~\eqref{eq:full_law_convergence}, and therefore their sum tends to zero.
	
	Let $f$ be any bounded observable recorded at the same resolution, and let $f_{jk}$ be its value in the bin $B_j\times C_k$. Using the expectations with respect to the two probability measures, we have
	\begin{align}\label{eq:expectation_from_full_law}
		\left|\E_{\mu_{T,Z_0}}[f]-\E_{\mu_{\rm st}}[f]\right|
		&=
		\left|\sum_{j,k}\left(P_{jk}(T)-Q_{jk}\right)f_{jk}\right|\nonumber\\
		&\leq
		\sup_{j,k}|f_{jk}|
		\sum_{j,k}\left|P_{jk}(T)-Q_{jk}\right|
		\longrightarrow0.
	\end{align}
	The first equality in Eq.~\eqref{eq:expectation_from_full_law} follows from the definition of the expectation value using the finite-resolution joint-bin probabilities $P_{jk}(T)$ and $Q_{jk}$. The second line follows from the triangle inequality and from the bound $|f_{jk}|\leq\sup_{j,k}|f_{jk}|$. The last limit then follows from Eq.~\eqref{eq:l1_from_full_law}. Hence relaxation of the full probability law already implies relaxation of every bounded expectation. Any marginal distribution converges for the same reason, because a marginal probability is obtained by summing the joint probabilities $P_{jk}(T)$ over a finite set of bins.
	
	The entropy follows just as directly. At the same finite resolution, Eq.~\eqref{eq:global_discrete_entropy} gives
	\begin{align}\label{eq:entropy_from_full_law}
		S_{\Delta\theta,\Delta\omega}[\rho_T]
		&=-\sum_{j,k}P_{jk}(T)\log P_{jk}(T)\nonumber\\
		&\longrightarrow
		-\sum_{j,k}Q_{jk}\log Q_{jk}
		=S_{\Delta\theta,\Delta\omega}[\rho_{\rm st}] .
	\end{align}
	The first equality in Eq.~\eqref{eq:entropy_from_full_law} is the definition of the finite-resolution entropy already used in Eq.~\eqref{eq:global_discrete_entropy}. To pass from the first line to the second, we use Eq.~\eqref{eq:full_law_convergence} and the continuity of $-x\log x$ on $[0,1]$, with $0\log0=0$. Since the number of bins is finite, the limit can be taken term by term inside the sum. The last equality is simply the same definition of the finite-resolution entropy applied to the stationary probabilities $Q_{jk}$.
	
	Thus, once the full recorded probability law relaxes, its marginals, all bounded expectations, and its finite-resolution entropy relax automatically.

	\section{Time reversal and forward/backward sampling}\label{sec:time_reversal}
	
	We now spell out the relation between predictions obtained by sampling the same deterministic trajectory forward and backward in time. This point is useful because time-reversal invariance of the dynamics does not mean that the two finite-$T$ probability distributions obtained from the same IC must be identical.
	
	Let $\Phi_t$ be an autonomous reversible dynamics and let $R$ be the time-reversal operation. By definition, $R$ is an involution, $R^2=1$, and
	\begin{equation}\label{eq:time_reversal_sm}
		R\circ\Phi_t\circ R=\Phi_{-t}.
	\end{equation}
	For a fixed IC $Z_0$, define the forward and backward occupation measures by
	\begin{equation}\label{eq:forward_backward_measures}
		\mu_{T,Z_0}(A)=\frac{1}{T}\int_0^T\mathbb{1}(\Phi_t(Z_0)\in A)\dd t,
		\qquad
		\mu_{-T,Z_0}(A)=\frac{1}{T}\int_0^T\mathbb{1}(\Phi_{-t}(Z_0)\in A)\dd t .
	\end{equation}
	The notation $\mu_{-T,Z_0}$ therefore means sampling the interval $[-T,0]$ while keeping $T>0$.
	
	Using Eq.~\eqref{eq:time_reversal_sm} in the second definition of Eq.~\eqref{eq:forward_backward_measures} gives
	\begin{align}
		\mu_{-T,Z_0}(A)
		&=\frac{1}{T}\int_0^T\mathbb{1}(R\Phi_t(RZ_0)\in A)\dd t \nonumber\\
		&=\frac{1}{T}\int_0^T\mathbb{1}(\Phi_t(RZ_0)\in R^{-1}A)\dd t \nonumber\\
		&=\mu_{T,RZ_0}(R^{-1}A).
		\label{eq:time_reversal_measure_relation}
	\end{align}
	In the first equality we replaced $\Phi_{-t}$ by $R\Phi_tR$ using Eq.~\eqref{eq:time_reversal_sm}. In the second equality we used the elementary equivalence $Rx\in A$ if and only if $x\in R^{-1}A$. The last equality is then precisely the definition of the forward occupation measure in Eq.~\eqref{eq:forward_backward_measures}, but starting from the reversed IC $RZ_0$. Thus, in general,
	\begin{equation}\label{eq:backward_not_forward}
		\mu_{-T,Z_0}\neq\mu_{T,Z_0};
	\end{equation}
	time reversal instead relates backward sampling from $Z_0$ to forward sampling from $RZ_0$.
	
	\emph{Finite measurement resolution.}
	The entropy used in the Letter refers to recorded outcomes, so we must also specify how the measurement bins transform. Let $\{B_\alpha\}$ be the finite partition of the recorded state space. We call this partition time-reversal symmetric when, for every bin $B_\alpha$, its image under $R$ is exactly another bin of the same partition. In formulas, there is a permutation $\pi$ of the bin labels such that
	\begin{equation}\label{eq:bin_permutation}
		R^{-1}B_\alpha=B_{\pi(\alpha)}.
	\end{equation}
	Writing $P^-_\alpha(T)=\mu_{-T,Z_0}(B_\alpha)$ and $P^+_\alpha(T;RZ_0)=\mu_{T,RZ_0}(B_\alpha)$, Eqs.~\eqref{eq:time_reversal_measure_relation} and \eqref{eq:bin_permutation} give
	\begin{equation}\label{eq:probability_permutation}
		P^-_\alpha(T)=P^+_{\pi(\alpha)}(T;RZ_0).
	\end{equation}
	Hence time reversal only relabels the probabilities. The finite-resolution Shannon entropy is therefore unchanged:
	\begin{align}
		-\sum_\alpha P^-_\alpha(T)\log P^-_\alpha(T)
		&=-\sum_\alpha P^+_{\pi(\alpha)}(T;RZ_0)
		\log P^+_{\pi(\alpha)}(T;RZ_0)\nonumber\\
		&=-\sum_\alpha P^+_\alpha(T;RZ_0)\log P^+_\alpha(T;RZ_0).
		\label{eq:entropy_time_reversal}
	\end{align}
	The second equality is only a relabeling of the finite sum: since $\pi$ is a permutation, every bin appears exactly once on both sides. For the optimal log-loss this gives
	\begin{equation}\label{eq:loss_time_reversal}
		L^*_{-T,Z_0}=L^*_{T,RZ_0}.
	\end{equation}
	Notice that Eq.~\eqref{eq:loss_time_reversal} does \emph{not} imply $L^*_{-T,Z_0}=L^*_{T,Z_0}$. Equality for the same IC requires the additional property that the forward distributions generated from $Z_0$ and $RZ_0$ have the same entropy.
	
	A simple mechanical example makes the meaning of Eq.~\eqref{eq:bin_permutation} transparent. For the usual time reversal $R(q,p)=(q,-p)$, take position bins $B_j$ and momentum bins $C_k$ arranged symmetrically about $p=0$. If $-C_k=C_{\bar k}$, then the joint bin $B_j\times C_k$ is mapped exactly to $B_j\times C_{\bar k}$. Time reversal has therefore done nothing but exchange the labels $k$ and $\bar k$. On the other hand, suppose that on the positive side one records a single momentum bin $C_+=[0,2\Delta p)$, while on the negative side the same interval is split into two bins $C^-_1=[-2\Delta p,-\Delta p)$ and $C^-_2=[-\Delta p,0)$. Then $R C_+=C^-_1\cup C^-_2$, rather than one recorded bin. A probability assigned to $C_+$ is split between two outcomes after time reversal, so the finite-resolution entropy need not be exactly preserved. This is why the statement about the entropy requires a time-reversal-symmetric measurement partition.
	
	\emph{The ring.}
	For the ring, $Z=(\theta,\omega)$ and $R(\theta,\omega)=(\theta,-\omega)$. With the IC $Z_0=(\theta_0,\omega_0)$ fixed, forward sampling traverses the incomplete arc $I_r^+(\theta_0)=\{\theta_0+2\pi u\ {\rm mod}\ 2\pi:0\leq u<r\}$, whereas backward sampling traverses $I_r^-(\theta_0)=\{\theta_0-2\pi u\ {\rm mod}\ 2\pi:0\leq u<r\}$. The two finite-$T$ densities are therefore generally different. They are related by the reflection $\theta\mapsto2\theta_0-\theta$ on the ring. This reflection has unit Jacobian and maps the ring onto itself, so the continuous finite part of the entropy used in Fig.~2 of the main text is the same in the two directions. Equivalently, if the finite angular bins are chosen symmetrically under this reflection, their discrete entropies are exactly equal. For an arbitrary fixed bin origin the equality is recovered in the resolution limit used in Sec.~3. Thus the regularized learning curve plotted in Fig.~2 satisfies $L^*_{-T,Z_0}=L^*_{T,Z_0}$, even though the two finite-$T$ densities need not coincide. When only $(\theta_0,E)$ is known, the two equally weighted signs of $\omega_0$ are exchanged by time reversal and the same conclusion holds for the entropy.
	
	\emph{Stationary limits and special ICs.}
	If the limits exist, let $\mu_{Z_0}^{\rm st,+}=\lim_{T\to\infty}\mu_{T,Z_0}$ and $\mu_{Z_0}^{\rm st,-}=\lim_{T\to\infty}\mu_{-T,Z_0}$. Taking $T\to\infty$ in Eq.~\eqref{eq:time_reversal_measure_relation} gives
	\begin{equation}\label{eq:stationary_time_reversal}
		\mu_{Z_0}^{\rm st,-}(A)
		=
		\mu_{RZ_0}^{\rm st,+}(R^{-1}A).
	\end{equation}
	Again, this does not force $\mu_{Z_0}^{\rm st,-}=\mu_{Z_0}^{\rm st,+}$ for the same IC. Special ICs can select special orbits for which the two limits differ. A simple possibility is a heteroclinic orbit: the trajectory approaches one invariant set as $t\to+\infty$ and a different invariant set as $t\to-\infty$. The forward occupation measure is then determined by the first asymptotic set, while the backward occupation measure is determined by the second. If instead the fixed IC selects a trajectory for which the two limits coincide, $\mu_{Z_0}^{\rm st,+}=\mu_{Z_0}^{\rm st,-}\equiv\mu_{Z_0}^{\rm st}$, then the forward and backward predictions converge to the same trajectory-selected stationary law and, at the same finite resolution, their optimal log-losses converge to the same plateau. The ring is of this latter type. No prior over ICs is involved anywhere in this discussion: all the measures above are selected by the fixed IC $Z_0$ through its deterministic trajectory.

	\bibliography{bibliography}